\documentclass[aps,prb,showpacs,twocolumn,amsmath,amssymb]{revtex4-2}

\usepackage{amsmath}    % need for subequations
\usepackage{amssymb}
\usepackage{graphicx}   % for figures
\usepackage{graphics}% Include figure files
\usepackage{dcolumn}% Align table columns on decimals point
\usepackage{bm}% bold math

\usepackage{multirow}

\usepackage{latexsym}
\usepackage{amssymb}
\usepackage{amsmath}

\usepackage{color}
\usepackage{epstopdf}

\usepackage[normalem]{ulem}

\begin{document}
\title{Elastic moduli of blue phases of cholesteric liquid crystals with low chirality}
\author{V.~A.~Chizhikov$^{\phantom{;}a,b}$}
\email{chizhikov@crys.ras.ru}
\author{V.~E.~Dmitrienko$^{\phantom{;}a,c}$}
\email{dmitrien@crys.ras.ru}
\address{$^a$Shubnikov Institute of Crystallography of the Kurchatov Complex of Crystallography and Photonics of the National Research Center ``Kurchatov Institute'', Moscow, 119333 Russia,}
\address{$^b$MIREA --- Russian Technological University (Institute of Radio Electronics and Informatics), Moscow, 119454 Russia,}
\address{$^c$Osipyan Institute of Solid State Physics of the Russian Academy of Sciences, Chernogolovka, Moscow Region, 142432 Russia}

\begin{abstract} %\Large
A new theoretical approach has been developed to describe the elastic properties of cubic blue phases of cholesteric liquid crystals (LCs). Blue phases are three-dimensional periodic chiral liquids with local anisotropy of the average orientation of molecules, and due to their periodicity, they have lattice elastic moduli characteristic of ordinary crystalline solids. The rigid tensor approximation, which works well at low chirality parameter ($\kappa\ll1$), was used to calculate the elastic moduli of the experimentally observed blue phases $O^8$ (BPI) and $O^2$ (BPII). It is shown that in the one-constant approximation for Frank moduli of LCs ($K_{11}=K_{22}=K_{33}$), the cubic lattice of blue phases has isotropic elasticity, and the Lam\'e's first parameter $\lambda_\mathrm{L}$ and Poisson's ratio $\nu$ are equal to zero. It is found that the sign of the Poisson's ratio is determined by the ratio of elastic moduli $K_0/K_1$; in particular, when $K_0>K_1$, the Poisson's ratio is negative.
\end{abstract}
\maketitle

%\Large

\section{Introduction}
\label{sec:intro}

Blue phases have been a popular subject of research in liquid crystal (LC) physics for many years (reviews of early work can be found in \cite{Belyakov1985,Wright1989,Crooker1989}, and a recent extensive development of this topic is presented in \cite{Oswald2005,Collings2014,Gleeson2015,Wu2022,Tubiana2024}). In terms of their structure, blue phases are self-organising three-dimensional periodic chiral fluids with local anisotropy of the average orientation of molecules, which is described by an order parameter in the form of a second-rank traceless tensor. Local anisotropy changes in a complex way within the unit cell, forming regions with double twist and inevitable defects (disclinations). The periodicity of the structure causes wavelength-dependent light diffraction, resulting in individual crystallites being colored differently, often blue, hence the name of these phases.

An attractive aspect of the theory of blue phases is that it is based on a relatively simple Landau--de~Gennes free energy functional containing a small number of parameters. This made it possible to construct a detailed picture of the structure and phase transitions of real phases and compare it with experiment \cite{Belyakov1985,Wright1989,Crooker1989,Brazovskii1975,Brazovskii1978,Grebel1983}. Similar expressions for the free energy are also used in other areas of physics, for example, in the theory of weak crystallization \cite{Kats1993} and to describe chiral magnets \cite{Stishov2023}, and therefore the theory of elastic properties developed in this article can find application both in these cases and for other spatially periodic chiral structures \cite{Dolganov2022,Baklanova2023}.

It is very important that, due to the periodicity, blue phases exhibit nonzero (albeit very small) rigidity \cite{Collings2014,Clark1984,Cladis1984,Kleiman1984}, inherent in all crystalline solids. This means that blue phases possess lattice elastic moduli, the calculation of which is the subject of this article. These calculations were performed within the framework of the same Landau--de~Gennes energy functional used to describe the structure and phase transitions.

Since the 1980s, a huge number of efforts have been devoted to the study of the diffraction optical properties of blue phases, for which they are often called photonic crystals. As in the case of X-ray diffraction in ordinary crystals, a detailed study of the spectral intensity and polarization of reflected and transmitted through the crystal light beams, and in particular circular dichroism, made it possible to thoroughly and quantitatively study the structure of all three blue phases BPI, BPII, BPIII, as well as the helical cholesteric phase \cite{Kizel1984, Demikhov1987, Demikhov1990, Demikhov1992}. In particular, it was shown \cite{Belyakov1985b} that the dependence of the structure of blue phases and cholesteric phase on temperature is determined by a single universal scalar order parameter --- the mean square of the tensor order parameter. A noticeable jump in this parameter is observed during the transition from an isotropic liquid to blue phases, while its jumps during the transitions between blue phases or into the cholesteric phase are very small, even though all transitions are first-order. This paper shows that the same behavior is characteristic of the elastic moduli of blue phases with low chirality.

It should be noted that, in addition to their nontrivial physical properties and potential practical applications, blue phases are also interesting as an example of systems with relatively simple local interactions at the intermolecular level that can self-organize into complex spatially heterogeneous macroscopic objects. It is possible that similar simple yet effective self-organization mechanisms act in living organisms today and were particularly important for the origin of life.

\section{Free energy of a cholesteric liquid crystal}
\label{sec:energy}

To study the structure and physical properties of the cholesteric LC, we will use the dimensionless Landau--de~Gennes free energy density $\varphi$ \cite{Belyakov1985,Wright1989,Collings2014}, written as a function of the order parameter $\hat{\chi}(\mathbf{r})$, which is a symmetric traceless tensor describing the local biaxial anisotropy of the cholesteric LC:
\begin{equation}
	\label{eq:varphi}
	\begin{array}l
		\varphi = \varphi_\mathrm{grad} + \varphi_\mathrm{bulk} , \vspace{0.2cm} \\ 
		\varphi_\mathrm{grad} = \kappa^2 \{\lVert \boldsymbol{\nabla} \times \hat{\chi} \pm \hat{\chi} \rVert^2 + \eta (\boldsymbol{\nabla} \cdot \hat{\chi})^2 \} , \vspace{0.2cm} \\
		\varphi_\mathrm{bulk} = \tau \, \mathrm{Tr}(\hat{\chi}^2) - \sqrt{6} \, \mathrm{Tr}(\hat{\chi}^3) + 2 \, \mathrm{Tr}(\hat{\chi}^4) .
	\end{array}
\end{equation}
Here $\varphi_\mathrm{grad}$ and $\varphi_\mathrm{bulk}$ are the gradient and bulk parts of the free energy density, respectively, and the mathematical notations included in $\varphi_\mathrm{grad}$ are defined as follows:
\[
(\boldsymbol{\nabla} \times \hat{A})_{\alpha\beta} = \epsilon_{\alpha\mu\nu} \nabla_\mu A_{\nu\beta} ,
\]
\[
(\boldsymbol{\nabla} \cdot \hat{A})_\alpha = \nabla_\mu A_{\mu\alpha} ,
\]
\[
\lVert \hat{A} \rVert = \sqrt{ A_{\alpha\beta}  A_{\alpha\beta} } .
\]
Here $\hat{\epsilon}$ is the antisymmetric Levi-Civita pseudotensor. The last expression is the definition of the Frobenius norm of a matrix. The meaning of the constants $\kappa$, $\eta$, $\tau$, included in the energy, will be explained in the text a little below.

Note that there is an alternative approach to the description of cholesteric LCs based on the order parameter in the form of the {\it director} $\mathbf{n}$ and the Frank--Oseen free energy
\begin{equation}
	\label{eq:Frank}
	\begin{array}c
		F = \tfrac12 K_{11} (\mathrm{div} \, \mathbf{n})^2 + \tfrac12 K_{22} (\mathbf{n} \cdot \mathrm{curl} \, \mathbf{n} \pm q)^2 \vspace{0.2cm} \\ 
		+ \tfrac12 K_{33} [\mathbf{n} \times \mathrm{curl} \, \mathbf{n}]^2 .
	\end{array}
\end{equation}
An important advantage of the approach based on the second-rank tensor $\hat{\chi}$ for studying the properties of blue phases is that, unlike the director field $\mathbf{n}(\mathbf{r})$, the field $\hat{\chi}(\mathbf{r})$ has no singularities at topological defects (disclinations). With regard to elasticity, this means that there is no need to take into account the contribution from defects separately from the bulk of the crystal. In addition, as will be shown below, a precondition for the stability of blue phases near the temperature of the transition from the isotropic to the cholesteric phase may be the biaxiality of the tensor $\hat{\chi}$, which is fundamentally absent in the director field. In the uniaxial case, the transition from (\ref{eq:varphi}) to (\ref{eq:Frank}) is accomplished by replacing
\[
\chi_{\alpha\beta} \sim n_\alpha n_\beta - \tfrac13 \delta_{\alpha\beta} .
\]

The gradient part $\varphi_\mathrm{grad}$ of the free energy density (\ref{eq:varphi}) is everywhere non-negative, since it contains only squares. This is achieved by including the positive term $\kappa^2\,\mathrm{Tr}(\hat{\chi}^2)$, which could be attributed to the bulk part of the energy by redefining the parameter $\tau$ \cite{Belyakov1985}. Note that although different ways of dividing the energy into conventional gradient and bulk parts cannot lead to different physical results, they can significantly affect the understanding of these results. In particular, the term $\kappa^2\,\mathrm{Tr}(\hat{\chi}^2)$ can compensate for the part of the gradient energy associated with the double twist of the order parameter field $\hat{\chi}(\mathbf{r})$. In this case, the energy gain from double twist will be taken into account in the bulk term $\varphi_\mathrm{bulk}$.

The number of free parameters in energy (\ref{eq:varphi}) is minimized. Here $\kappa$ is a dimensionless positive parameter called {\it chirality}, which characterizes the twist strength of the order parameter field $\hat{\chi}(\mathbf{r})$. The sign of the helical twist depends on the choice of plus or minus in the first term of the gradient energy. In a physical sense, chirality $\kappa$ is determined by the ratio of two characteristic lengths:
\begin{equation}
	\label{eq:kappa}
	\kappa = 2 \pi \xi / p .
\end{equation}
Here $\xi$ is the correlation length, which is close in magnitude to the length of the organic molecules that make up the LC (10--50\AA), and $p$ is the pitch of the cholesteric helix, comparable in size to the wavelength of the visible spectrum of light (10$^3$--10$^4$\AA). Thus, $\kappa\ll1$ for most cholesteric LCs.

Note that equation~(\ref{eq:varphi}) uses dimensionless coordinates $\mathbf{r}^\prime=4\pi \mathbf{r}/p$, so the expression for $\varphi_\mathrm{grad}$ is not applicable in the nematic limit $p\rightarrow\infty$. Using the coordinate renormalization $\mathbf{r}^{\prime\prime}=2\mathbf{r}/\xi$ in the case of a non-chiral LC, we arrive at the expression for the gradient energy density
\[
\varphi_\mathrm{grad} = \lVert \boldsymbol{\nabla} \times \hat{\chi} \rVert^2 + \eta (\boldsymbol{\nabla} \cdot \hat{\chi})^2 .
\]

The parameter $\eta$ is equal to the ratio of the elastic moduli of the liquid crystal: $\eta=K_0/K_1$. The case $\eta=1$ ($K_0=K_1$) is called {\it one-constant}. It should be noted that the Frank--Oseen energy (\ref{eq:Frank}), often used to describe nematic and cholesteric liquid crystals, contains three bulk elastic moduli $K_{11}, K_{22}, K_{33}$. The Frank moduli are related to $K_0$ and $K_1$ as follows:
\begin{equation}
	\label{eq:Frank-moduli}
	K_{11} = K_{33} \sim \tfrac12 (K_0 + K_1) , \phantom{x} K_{22} \sim K_1 .
\end{equation}
Thus, in the one-constant approximation, all Frank moduli are equal to each other, $K_{11}=K_{22}=K_{33}$.

The last free parameter $\tau$ plays the role of temperature. Since the gradient energy density $\varphi_\mathrm{grad}$ is everywhere non-negative, it is completely minimized by the field $\hat{\chi}(\mathbf{r})=0$, which corresponds to an isotropic liquid. For $\tau\geqslant0.25$, this solution also corresponds to a minimum of the bulk energy $\varphi_\mathrm{bulk}$. For $\tau<0.25$, the bulk energy is minimized by the uniaxial order parameter, as in the homogeneous nematic phase
\begin{equation}
	\label{eq:chiN}
	\hat{\chi}_\mathrm{N} \sim \left(
	\begin{array}{rrr}
		\tfrac23 & 0 & 0 \\
		0 & -\tfrac13 & 0 \\
		0 & 0 & -\tfrac13
	\end{array}
	\right)
\end{equation}
with the director oriented along the $x$ axis. In principle, simultaneous minimization of gradient and bulk energies can be maintained under the following conditions:
\begin{equation}
	\label{eq:varphigrad-minimum}
	\left\{
	\begin{array}l
		\boldsymbol{\nabla} \times \hat{\chi} \pm \hat{\chi} = 0 , \vspace{0.2cm} \\
		\boldsymbol{\nabla} \cdot \hat{\chi} = 0 .
	\end{array}
	\right.
\end{equation}
Non-trivial solutions of the system (\ref{eq:varphigrad-minimum}) have the form of biaxial helices of the type
\begin{equation}
	\label{eq:chiHelix}
	\hat{\chi}_\mathrm{helix} \sim \left(
	\begin{array}{ccc}
		\cos z & \pm \sin z & 0 \\
		\pm \sin z & -\cos z & 0 \\
		0 & 0 & 0
	\end{array}
	\right) ,
\end{equation}
where the sign of the sines corresponds to a right-handed (plus) or left-handed (minus) helix. The fact that the bulk energy is minimized by a uniaxial order parameter, and the gradient energy is minimized by a biaxial one, leads to an interesting competition between these two energies at $\tau\approx0$, which gives rise to the blue phases of LCs. As the temperature decreases, the bulk energy wins, and the cholesteric LC passes into a uniaxial helical phase with the order parameter
\begin{equation}
	\label{eq:chiCh}
	\hat{\chi}_\mathrm{Ch} \sim \left(
	\begin{array}{ccr}
		\tfrac13 + \cos z & \pm \sin z & 0 \\
		\pm \sin z & \tfrac13 - \cos z & 0 \\
		0 & 0 & -\tfrac23
	\end{array}
	\right) ,
\end{equation}
which can be described using a director rotating in the $xy$ plane when the $z$ coordinate changes.

To solve the problem of minimizing the free energy (\ref{eq:varphi}) of the crystalline blue phase, it is convenient to go to the Fourier representation of the order parameter:
\begin{equation}
	\label{eq:chi-fourier}
	\hat{\chi}(\mathbf{r}) = \sum_\mathbf{k} \hat{\chi}_\mathbf{k} \exp(i \mathbf{k} \cdot \mathbf{r}) ,
\end{equation}
where the summation is over the vectors $\mathbf{k}$ of the reciprocal lattice, $\hat{\chi}_{-\mathbf{k}}=\hat{\chi}_\mathbf{k}^\ast$ due to the reality of the field $\hat{\chi}(\mathbf{r})$. Then one can easily find the average free energy density
\begin{equation}
	\label{eq:phigrad-fourier}
	\begin{array}{ll}
		\left< \varphi_\mathrm{grad} \right> = & \kappa^2 \sum\limits_\mathbf{k} \{ (k^2 + 1) \, \mathrm{Tr}(\hat{\chi}_\mathbf{k} \cdot \hat{\chi}^\ast_\mathbf{k}) \vspace{0.2cm} \\ 
		& + (\eta - 1) \mathbf{k} \cdot \hat{\chi}_\mathbf{k} \cdot \hat{\chi}^\ast_\mathbf{k} \cdot \mathbf{k} \vspace{0.2cm} \\ 
		& \pm 2 i \, \mathrm{Tr}(\mathbf{k} \times (\hat{\chi}_\mathbf{k} \cdot \hat{\chi}^\ast_\mathbf{k})) \} ,
	\end{array}
\end{equation}
\begin{equation}
	\label{eq:phibulk-fourier}
	\begin{array}{ll}
		\left< \varphi_\mathrm{bulk} \right> = & \tau \sum\limits_{\Sigma\mathbf{k}_i=0} \mathrm{Tr}(\hat{\chi}_{\mathbf{k}_1} \cdot \hat{\chi}_{\mathbf{k}_2}) \vspace{0.2cm} \\
		&  - \sqrt{6} \sum\limits_{\Sigma\mathbf{k}_i=0} \mathrm{Tr}(\hat{\chi}_{\mathbf{k}_1} \cdot \hat{\chi}_{\mathbf{k}_2} \cdot \hat{\chi}_{\mathbf{k}_3})  \vspace{0.2cm} \\
		& + 2 \sum\limits_{\Sigma\mathbf{k}_i=0} \mathrm{Tr}(\hat{\chi}_{\mathbf{k}_1} \cdot \hat{\chi}_{\mathbf{k}_2} \cdot \hat{\chi}_{\mathbf{k}_3} \cdot \hat{\chi}_{\mathbf{k}_4}) .
	\end{array}
\end{equation}
The cross product of vector and matrix in (\ref{eq:phigrad-fourier}) has the same meaning as in (\ref{eq:varphi}). In (\ref{eq:phibulk-fourier}), the sums are taken over two, three, and four reciprocal lattice vectors, respectively, provided that the sum of these vectors is zero.

\section{Structure of blue phases}
\label{sec:structure}

As can be seen from the expressions (\ref{eq:phigrad-fourier}), (\ref{eq:phibulk-fourier}), the average free energy density is a function of two types of variables: the components of the wave vectors $\mathbf{k}$ and the components of the corresponding tensors $\hat{\chi}_\mathbf{k}$,
\[
\left< \varphi \right> = f(\{\mathbf{k}\}, \{\hat{\chi}_\mathbf{k}\}) .
\]
Since the number of variables of the function $f$ is in principle undefined, and should ideally be infinite, the search for structures that minimize the free energy is, generally speaking, a nontrivial problem. Experimental data (observed Bragg reflections) are used to limit the number of Fourier components $\hat{\chi}_\mathbf{k}$ taken into account \cite{Belyakov1985,Wright1989,Collings2014}. Significant restrictions on the tensor form of $\hat{\chi}_\mathbf{k}$ arise from taking into account the symmetry of the blue phases \cite{Belyakov1985}. For example, it is known that the blue phases BPI and BPII have cubic symmetry. Of course, there is no guarantee that the structures found in modelling correspond to the global minimum of free energy. Nevertheless, as a result of intensive research over many years, several cubic phases have been mathematically described that have better energy at $\tau\approx0$ than the helical cholesteric phase (\ref{eq:chiCh}). Among them is the phase with the space group $O^5$ ($I432$), which we analyzed in detail in \cite{Chizhikov2024}, as well as phases with the symmetries $O^8$ ($I4_132$) and $O^2$ ($P4_232$), associated with the experimentally observed phases BPI and BPII, respectively,  for which an attempt was previously made to construct a theory of elasticity \cite{Dmitrienko1986}.

In this paper, we consider almost ideal structures of blue phases. Here, {\it almost} means that the structure corresponds to a local minimum of the gradient part of the free energy, but does not minimize the total energy. As already mentioned, the average gradient energy density $\left<\varphi_\mathrm{grad}\right>$ has a minimum if the tensor field of the order parameter consists of Fourier harmonics of the form
\begin{equation}
	\label{eq:expikr}
	\hat{\chi}_\mathbf{k} \exp(i \mathbf{k} \cdot \mathbf{r}) ,
\end{equation}
\begin{equation}
	\label{eq:chiLR}
	\hat{\chi}_\mathbf{k} = \tfrac12 A (\mathbf{m}_1 \mp i \mathbf{m}_2) \otimes (\mathbf{m}_1 \mp i \mathbf{m}_2) ,
\end{equation}
representing biaxial helices. Here, vectors $\mathbf{m}_1$ and $\mathbf{m}_2$ form an orthonormal basis in the plane perpendicular to the wave vector $\mathbf{k}$,
\begin{equation}
	\label{eq:m1m2}
	[\mathbf{m}_1 \times \mathbf{m}_2] = \mathbf{k} / |\mathbf{k}| \equiv \mathbf{n}_\mathbf{k} .
\end{equation}
For example, the set of vectors $\mathbf{m}_1=(100)$, $\mathbf{m}_2=(010)$, $\mathbf{k}=(001)$ defines a Fourier harmonic, which, when combined with its complex conjugate, forms a helix (\ref{eq:chiHelix}). The factor $\tfrac12$ provides a convenient normalization:
\begin{equation}
	\label{eq:chinorm}
	\lVert \hat{\chi}_\mathbf{k} \rVert^2 = \mathrm{Tr} (\hat{\chi}_\mathbf{k} \cdot \hat{\chi}^\ast_\mathbf{k}) =  A^2 ,
\end{equation}
and the choice of the plus or minus sign determines whether the helix is left- or right-handed, respectively. The structure always consists of helices of the same chirality, determined by the properties of the cholesteric LC, which, in turn, are related to the chirality of its constituent molecules. Note that the absolute minimum of $\left<\varphi_\mathrm{grad}\right>$ is achieved for wave vectors of unit length, $k=1$. However, if the blue phase is composed of several symmetrically unrelated harmonics, then their wavenumbers $k$ can differ, and the equilibrium period of the lattice is determined by the competition of harmonics.

Since the blue phases possess high symmetry, most Fourier harmonics are related to each other by space group transformations. So, it is sufficient to define only the minimum number of independent harmonics for each phase and then generate the remaining harmonics using the group elements. For example, if the space group includes symmetry transformation
\begin{equation}
	\label{eq:symtransform}
	\underline{\mathbf{r}} = R \cdot \mathbf{r} + \mathbf{t} ,
\end{equation}
where $R$ is the rotation matrix and $\mathbf{t}$ is the translation (for non-symmorphic groups), then the Fourier harmonic (\ref{eq:expikr}) under the action of this symmetry element transforms into an equivalent harmonic
\begin{equation}
	\label{eq:expikr-und}
	\hat{\chi}_{\underline{\mathbf{k}}} \exp(i \underline{\mathbf{k}} \cdot \mathbf{r}) ,
\end{equation}
with
\begin{equation}
	\label{eq:k-und}
	\underline{\mathbf{k}} = R \cdot \mathbf{k} ,
\end{equation}
\begin{equation}
	\label{eq:chi-und}
	\hat{\chi}_{\underline{\mathbf{k}}} = \exp(-i \underline{\mathbf{k}} \cdot \mathbf{t}) \left[ R \cdot \hat{\chi}_\mathbf{k} \cdot R^{-1} \right] .
\end{equation}

The reciprocal lattice vector can be parallel or perpendicular to a rotation axis of the crystal. In the first case, upon rotation around the axis, $\underline{\mathbf{k}}=\mathbf{k}$ and, consequently, $\hat{\chi}_{\underline{\mathbf{k}}}=\hat{\chi}_\mathbf{k}$. In the second case, if the axis is 2-, 4-, or 6-fold, then there is a rotation through the angle $\pi$, for which $\underline{\mathbf{k}}=-\mathbf{k}$ and $\hat{\chi}_{\underline{\mathbf{k}}}=\hat{\chi}_\mathbf{k}^\ast$. In both cases, the condition (\ref{eq:chi-und}) imposes a strict constraint on the phase of the Fourier harmonic. In other cases, no symmetry constraints are imposed on the harmonic phase and it is chosen so as to minimize the bulk free energy $\left<\varphi_\mathrm{bulk}\right>$.

As already mentioned, blue phases in cholesteric LCs arise due to the competition between the gradient and bulk contributions to the free energy. Clearly, the gradient energy only loses out when the spatial lattice emerges. Indeed, the average gradient energy density $\left<\varphi_\mathrm{grad}\right>$ is minimized by biaxial helices (\ref{eq:expikr}), (\ref{eq:chiLR}) with wavenumber $k=1$. However, after the lattice emerges, even if the new structure is formed exclusively by biaxial helices, reciprocal lattice vectors of different lengths will inevitably emerge. For Fourier harmonics with $k\neq1$, the gradient energy is strictly positive:
\begin{equation}
	\label{eq:phihelix}
	\left<\varphi_\mathrm{grad}\right> \propto (k - 1)^2 > 0 .
\end{equation}
For the blue phase to be stable, the increase in the gradient energy must be compensated by a decrease in the bulk energy. At $\tau\geqslant0$, the only non-positive term in the expression (\ref{eq:varphi}) for the free energy density is $-\sqrt{6}\,\mathrm{Tr}(\hat{\chi}^3)$. This term reaches its smallest value when the order papameter becomes uniaxial. Thus, the existence of blue phases at $\tau\approx0$ is a consequence of the system's tendency toward maximum uniaxiality, that is, toward a state described by the director. For $\tau$ significantly less than zero, uniaxiality ultimately prevails, leading to a transition to the helical phase (\ref{eq:chiCh}). 

Let us describe the independent harmonics corresponding to the strongest Bragg reflections in the ideal blue phases $O^5$, $O^8$, and $O^2$. Here we restrict ourselves to structures consisting of left-handed helices (the choice of the plus sign in (\ref{eq:chiLR}) and the minus sign in (\ref{eq:varphi}) and (\ref{eq:phigrad-fourier})).

\subsection*{$O^5$}

The ideal structure of the blue phase $O^5$ ($I432$) consists of identical helices with axes along equivalent crystallographic directions $\left<110\right>$. The wave vectors $\mathbf{k}$ are parallel to the 2-fold axes and perpendicular to the 4-fold axes of the crystal point group, which, in accordance with the above, means that the phases of the helices are determined by symmetry (up to a sign; see the table on p.~385 in \cite{Belyakov1985}). Choosing $\mathbf{m}_1=(001)$, $\mathbf{m}_2=\tfrac1{\sqrt{2}}(1\bar{1}0)$ for the reciprocal lattice vector $\mathbf{k}\parallel[110]$, from (\ref{eq:chiLR}) we find 
\begin{equation}
	\label{eq:chi110-O5}
	\hat{\chi}_{110} = \frac{\chi_{\left<110\right>}}4 \left(
	\begin{array}{rrr}
		-1 & 1 & i\sqrt{2} \\
		1 & -1 & -i\sqrt{2} \\
		i\sqrt{2} & -i\sqrt{2} & 2
	\end{array}
	\right) .
\end{equation}
Here, instead of the index $\mathbf{k}$ for the Fourier harmonic, it is convenient to use the Miller indices $hk\ell$; $\mathbf{k}=\frac{2\pi}a(hk\ell)$, where $a$ is the cubic unit cell parameter. The remaining eleven harmonics are found by the formulas (\ref{eq:k-und}), (\ref{eq:chi-und}) with rotation elements from the point group $432$ ($\mathbf{t}=0$, since $I432$ is a symmorphic group).

The multiplier $\chi_{\left<110\right>}$ common to all harmonics is determined from the condition of minimum free energy. Here we have introduced the notation $\chi_{\langle hk\ell\rangle}$ for the real normalization factor,
\[
\chi_{\langle hk\ell\rangle} = \pm \lVert \hat{\chi}_{hk\ell} \rVert .
\]
As is common in crystal physics, the Miller indices $\langle hk\ell\rangle$ enclosed in angle brackets denote the set of reciprocal lattice vectors related with $hk\ell$ by transformations of the point group, in this case $O_h$ ($m\bar{3}m$). The inversion must be added here, since in general the reciprocal lattice vector $-\mathbf{k}$ may not be related to the vector $\mathbf{k}$ by an element of the group $O$ ($432$). Thus, the set $\langle hk\ell\rangle$ contains all vectors obtained from $hk\ell$ by permuting the coordinates and/or changing their signs. For general Miller indices, this set will contain 48 vectors, but in special cases the number of vectors may be smaller. For example, for Miller indices 110, there will be a total of $N_{\langle 110\rangle}=12$ vectors:
\[
\begin{array}l
	\langle 110\rangle = \{ 011, 0\bar{1}1, 01\bar{1}, 0\bar{1}\bar{1}, 101, 10\bar{1}, \\
	\phantom{\langle 110\rangle = \{} \bar{1}01, \bar{1}0\bar{1}, 110, \bar{1}10, 1\bar{1}0, \bar{1}\bar{1}0 \} .
\end{array}
\]
Due to symmetry, the normalization factor $\chi_{\langle hk\ell\rangle}$ must be the same for all equivalent harmonics. From now on, we will also use angle brackets to denote averaging, which we hope will not lead to much confusion.

One interesting fact is worth noting. Since the ideal $O^5$ phase consists of biaxial helices with wavenumber $k=1$, its average gradient energy is zero, $\left<\varphi_\mathrm{grad}\right>=0$. And since $\varphi_\mathrm{grad}$ is non-negative, it is also zero everywhere inside the unit cell, including both nearly uniaxial (director-described) regions and essentially biaxial regions (disclination cores). On the other hand, the blue phases $O^8$ and $O^2$, whose structures are described below, contain helices with different $k$, and hence their average gradient energy is strictly positive. Then the gradient energy density determined by the superposition of crossed biaxial helices turns out to be unevenly distributed over the unit cell. The value of $\varphi_\mathrm{grad}$ reaches a minimum in regions with double twist of the director field and a maximum at the disclination cores. This seems to be consistent with the generally accepted view that the gain in gradient energy is due to the double twist of the director field. However, the case of the $O^5$ phase shows that in the theory based on the tensor order parameter $\hat{\chi}(\mathbf{r})$ and the expression (\ref{eq:varphi}) for the free energy, the decrease in the gradient energy is not due to a double twist at all, but to biaxial helicess of the form (\ref{eq:chiLR}).

\begin{figure}[h]
	\begin{center}
		\includegraphics[width=8cm]{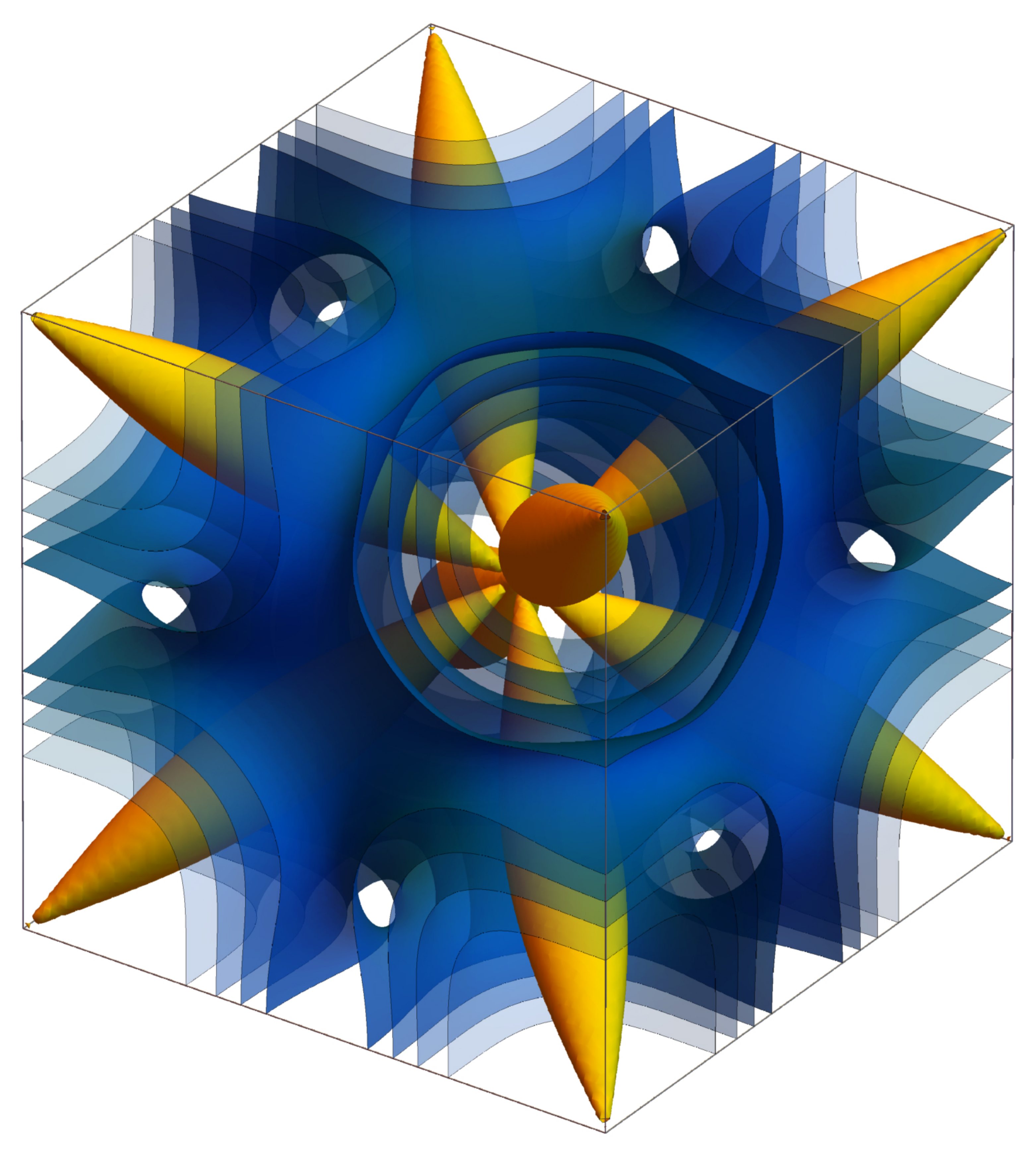}
		\caption{\label{fig:O5} Distribution of the value of $\mathrm{Tr}(\hat{\chi}^3)$, which determines the free energy gain, in the unit cell of the ideal blue phase $O^5$. The condition $\mathrm{Tr}(\hat{\chi}^3)=0$ (yellow surface) defines the boundary of topological defects (disclination cores). Light blue surfaces correspond to $\mathrm{Tr}(\hat{\chi}^3)$ values equal to 20, 40, 60, and 80\% of the maximum.}
	\end{center}
\end{figure}

The energy balance of the $O^5$ phase is largely determined by the term $-\sqrt{6}\,\mathrm{Tr}(\hat{\chi}^3)$ in the free energy density. Fig.~\ref{fig:O5} shows the $\mathrm{Tr}(\hat{\chi}^3)$ distribution in the unit cell of the ideal blue phase $O^5$.

\subsection*{$O^8$}

In the ideal structure of the $O^8$ ($I4_132$) phase, three main nonequivalent Fourier harmonics with Miller indices $\left<200\right>$, $\left<110\right>$, $\left<211\right>$ were experimentally detected. The phases of the harmonics are determined by symmetry, as in the $O^5$ case. Choosing $\mathbf{m}_1=(010)$, $\mathbf{m}_2=(001)$ for the reciprocal lattice vector $\mathbf{k}=\frac{2\pi}a(200)$, we obtain
\begin{equation}
	\label{eq:chi200-O8}
	\hat{\chi}_{200} = \frac{\chi_{\left<200\right>}}2 \left(
	\begin{array}{rrr}
		0 & 0 & 0 \\
		0 & 1 & i \\
		0 & i & -1
	\end{array}
	\right) ,
\end{equation}
and choosing $\mathbf{m}_1=\tfrac12(1,-1,\sqrt{2})$, $\mathbf{m}_2=\tfrac12(1,-1,-\sqrt{2})$ for $\mathbf{k}=\frac{2\pi}a(110)$, we have
\begin{equation}
	\label{eq:chi110-O8}
	\hat{\chi}_{110} = \frac{\chi_{\left<110\right>}}4 \left(
	\begin{array}{rrr}
		i & -i & \sqrt{2} \\
		-i & i & -\sqrt{2} \\
		\sqrt{2} & -\sqrt{2} & -2i
	\end{array}
	\right) .
\end{equation}
Note that the phase of the Fourier harmonic 110 differs for the $O^5$ and $O^8$ groups. This is due to the non-symmorphism of the space group $O^8$. For example, the rotation by $\pi$ about the $[001]$ axis, transforming the vector $\mathbf{k}=\frac{2\pi}a(110)$ into $-\mathbf{k}$, is accompanied by a shift $\mathbf{t}=\tfrac{a}2(101)$, which changes the sign of the right-hand side of the equation (\ref{eq:chi-und}).

Choosing $\mathbf{m}_1=\tfrac1{\sqrt{3}}(1\bar{1}\bar{1})$, $\mathbf{m}_2=\tfrac1{\sqrt{2}}(01\bar{1})$ for the wave vector $\mathbf{k}=\frac{2\pi}a(211)$, we obtain
\begin{equation}
	\label{eq:chi211-O8}
	\begin{array}{ll} 
		\hat{\chi}_{211} = \displaystyle \frac{\chi_{\left<211\right>}}{12} & \left[ \left(
		\begin{array}{rrr}
			2 & -2 & -2 \\
			-2 & -1 & 5 \\
			-2 & 5 & -1
		\end{array}
		\right) \right. \vspace{0.2cm} \\
		& + i\sqrt{6} \left. \left(
		\begin{array}{rrr}
			0 & 1 & -1 \\
			1 & -2 & 0 \\
			-1 & 0 & 2
		\end{array}
		\right) \right] .
	\end{array}
\end{equation}
The remaining Fourier harmonics can be found from the harmonics (\ref{eq:chi200-O8})--(\ref{eq:chi211-O8}) using formulas (\ref{eq:k-und}), (\ref{eq:chi-und}) with the elements of the space group $O^8$.

\begin{figure}[h]
	\begin{center}
		\includegraphics[width=8cm]{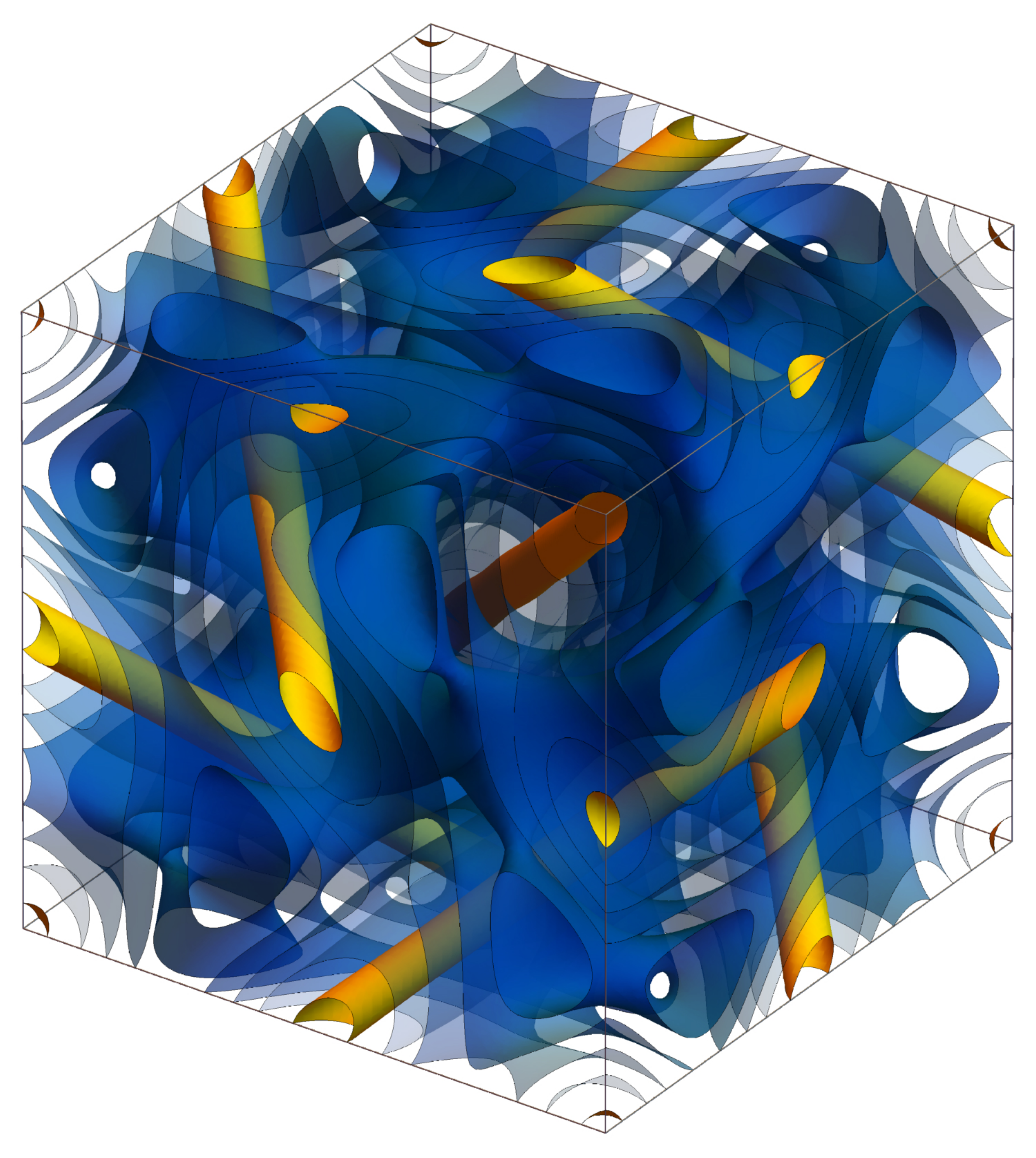}
		\caption{\label{fig:O8} Distribution of the value of $\mathrm{Tr}(\hat{\chi}^3)$, which determines the free energy gain, in the unit cell of the ideal blue phase $O^8$ (BPI). The condition $\mathrm{Tr}(\hat{\chi}^3)=0$ (yellow surface) defines the boundary of topological defects (disclination cores). Light blue surfaces correspond to $\mathrm{Tr}(\hat{\chi}^3)$ values equal to 20, 40, 60, and 80\% of the maximum.}
	\end{center}
\end{figure}

Fig.~\ref{fig:O8} shows the $\mathrm{Tr}(\hat{\chi}^3)$ distribution in the unit cell of the ideal blue phase $O^8$.

\subsection*{$O^2$}

The ideal structure of the blue phase $O^2$ ($P4_232$) contains two types of main Fourier harmonics: $\left<100\right>$ and $\left<110\right>$, whose phases are determined by symmetry. The 100 harmonic is similar to the 200 harmonic of the phase $O^8$, but with double period, see (\ref{eq:chi200-O8}):
\begin{equation}
	\label{eq:chi100-O2}
	\hat{\chi}_{100} = \frac{\chi_{\left<100\right>}}2 \left(
	\begin{array}{rrr}
		0 & 0 & 0 \\
		0 & 1 & i \\
		0 & i & -1
	\end{array}
	\right) ,
\end{equation}
while the 110 harmonic is the same as in the phase $O^5$ except for the value of the coefficient $\chi_{\left<110\right>}$, see (\ref{eq:chi110-O5}). The remaining Fourier harmonics can be found from these two using formulas (\ref{eq:k-und}), (\ref{eq:chi-und}) with elements of the space group $O^2$.

\begin{figure}[h]
	\begin{center}
		\includegraphics[width=8cm]{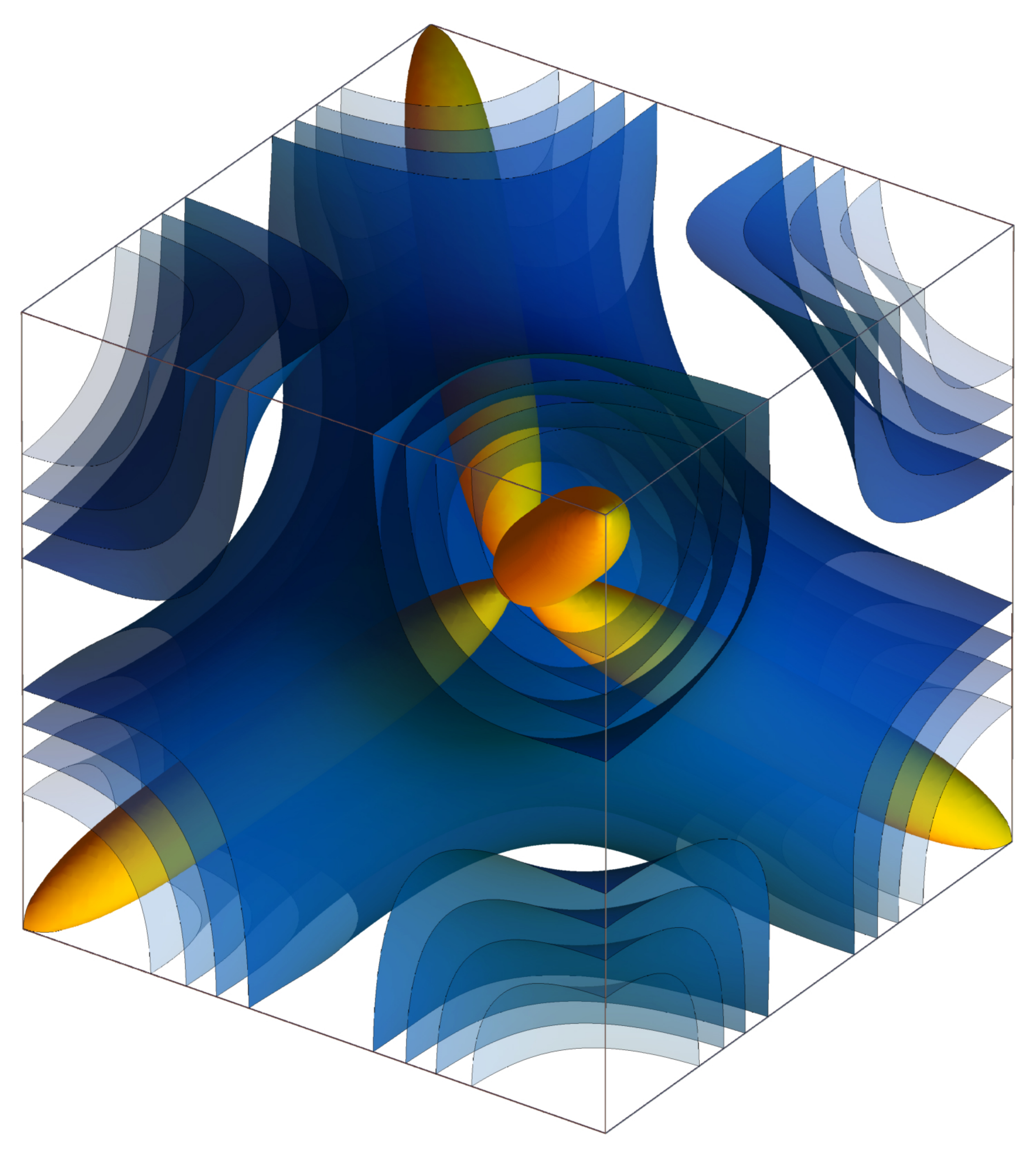}
		\caption{\label{fig:O2}  Distribution of the value of $\mathrm{Tr}(\hat{\chi}^3)$, which determines the free energy gain, in the unit cell of the ideal blue phase $O^2$ (BPII). The condition $\mathrm{Tr}(\hat{\chi}^3)=0$ (yellow surface) defines the boundary of topological defects (disclination cores). Light blue surfaces correspond to $\mathrm{Tr}(\hat{\chi}^3)$ values equal to 20, 40, 60, and 80\% of the maximum.}
	\end{center}
\end{figure}

Fig.~\ref{fig:O2} shows the $\mathrm{Tr}(\hat{\chi}^3)$ distribution in the unit cell of the ideal blue phase $O^2$.

\section{Elastic deformation of the cubic crystal lattice}
\label{sec:deform}

As already mentioned, the average free energy density $\left<\varphi\right>$ at given physical parameters $\kappa$, $\eta$, $\tau$ is determined by the sets of variables $\{\mathbf{k}\}$ and $\{\hat{\chi}_\mathbf{k}\}$. Suppose that the system is in a state of equilibrium corresponding to a local energy minimum. If mechanical stress is applied to the system, the increase in free energy caused by deformation will be a quadratic form in the increments of the variables:
\begin{equation}
	\label{eq:phidef1}
	\begin{array}{ll}
		\varphi_\mathrm{def}  = & \displaystyle \frac12 \frac{\partial^2 \left<\varphi\right>}{\partial k_i \partial k_j} \Delta k_i \Delta k_j \vspace{0.2cm} \\
		& \displaystyle + \frac{\partial^2 \left<\varphi\right>}{\partial k_i \partial \chi_j} \Delta k_i \Delta \chi_j \vspace{0.2cm} \\
		& \displaystyle + \frac12 \frac{\partial^2 \left<\varphi\right>}{\partial \chi_i \partial \chi_j} \Delta \chi_i \Delta \chi_j .
	\end{array}
\end{equation}
Here, indices $i$ and $j$ number all components of wave vectors $\mathbf{k}$ and tensors $\hat{\chi}_\mathbf{k}$; repeated indices imply summation. The increments $\Delta k_i$ of the components of the reciprocal lattice vectors are completely determined by the deformation tensor, while the changes $\Delta \chi_i$ of the components of the order parameter tensors arise due to the relaxation of the system to a new state of equilibrium:
\begin{equation}
	\label{eq:phidef-eq}
	\frac{\partial \varphi_\mathrm{def}}{\partial \Delta \chi_i} = \frac{\partial^2 \left<\varphi\right>}{\partial \chi_i \partial k_j} \Delta k_j +  \frac{\partial^2 \left<\varphi\right>}{\partial \chi_i \partial \chi_j} \Delta \chi_j = 0 .
\end{equation}
Using (\ref{eq:phidef-eq}), the elastic deformation energy of LC takes the form
\begin{equation}
	\label{eq:phidef2}
	\varphi_\mathrm{def}  = \frac12 \frac{\partial^2 \left<\varphi\right>}{\partial k_i \partial k_j} \Delta k_i \Delta k_j - \frac12 \frac{\partial^2 \left<\varphi\right>}{\partial \chi_i \partial \chi_j} \Delta \chi_i \Delta \chi_j .
\end{equation}

In \cite{Chizhikov2024}, it was shown that in cholesteric LCs with a small chirality parameter ($\kappa\ll1$), the distortion of tensors $\hat{\chi}_\mathbf{k}$ can be neglected during blue phase deformations: the so-called {\it rigid tensor approximation}. The second term discarded in (\ref{eq:phidef2}) has an order of $\kappa^4$. Since the explicit dependence on the vectors $\mathbf{k}$ of the reciprocal lattice is contained only in the gradient part of the free energy (\ref{eq:phigrad-fourier}), in this approximation the elastic deformation energy can be written as
\begin{equation}
	\label{eq:phidef3}
	\varphi_\mathrm{def} = \frac12 \sum_{\mathbf{k}} \frac{\partial^2 \left<\varphi_\mathrm{grad}\right>}{\partial k_\alpha \partial k_\beta} (k^\prime_\alpha - k_\alpha) (k^\prime_\beta - k_\beta) ,
\end{equation}
where $\mathbf{k}^\prime$ is the reciprocal lattice vector after deformation. The distorted vector $\mathbf{k}^\prime$ is related to the initial $\mathbf{k}$ in such a way that the phase of the Fourier harmonic at the displaced point $\mathbf{r}^\prime$ is preserved:
\begin{equation}
	\label{eq:phase}
	\mathbf{k}^\prime \cdot \mathbf{r}^\prime = \mathbf{k} \cdot \mathbf{r} .
\end{equation}
Under uniform deformation, the displacement of crystal points is determined by an affine transformation
\begin{equation}
	\label{eq:rdeform}
	\mathbf{r}^\prime = (1 + \hat{u}) \cdot \mathbf{r} ,
\end{equation}
with
\begin{equation}
	\label{eq:u}
	(\hat{u})_{\alpha\beta} = \frac{\partial (r^\prime_\alpha - r_\alpha)}{\partial r_\beta} .
\end{equation}
From (\ref{eq:phase}) it follows that
\begin{equation}
	\label{eq:kdeform}
	\mathbf{k}^\prime =  \mathbf{k} \cdot (1 + \hat{u})^{-1} .
\end{equation}

The tensor $(1 + \hat{u})$ may have an antisymmetric part that corresponds to the rotation of the crystal as a whole. In isotropic space, the energy does not change under rotations, and it is convenient to eliminate the rotation by symmetrizing the tensor $\hat{u}$. According to the polar decomposition theorem, there exists an orthogonal matrix $R$ such that
\[
(1 + \hat{u}) = R \cdot (1 + \hat{u}_\mathrm{s}) ,
\]
where $\hat{u}_\mathrm{s}$ is a symmetric matrix. Applying the rotation $R^{-1}$ to the deformed system, we obtain
\begin{equation}
	\label{eq:rdeform2}
	\mathbf{r}^\prime_\mathrm{new} = R^{-1} \cdot \mathbf{r}^\prime_\mathrm{old} = (1 + \hat{u}_\mathrm{s}) \cdot \mathbf{r} ,
\end{equation}
thereby achieving the desired result.

Let us now introduce the deformation tensor $\hat{\pi}$ (from the Greek $\pi\alpha\rho\alpha\mu\acute{o}\rho\phi\omega\sigma\eta$ --- deformation):
\begin{equation}
	\label{eq:pitensor}
	\hat{\pi} = \frac{\hat{u}_\mathrm{s}}{1 + \hat{u}_\mathrm{s}} ,
\end{equation}
which, by definition, is also symmetric. The tensor $\hat{\pi}$ is more convenient for our problem than the commonly used Cauchy--Green deformation tensor $\hat{\varepsilon}$, and is related to the latter as follows:
\begin{equation}
	\label{eq:pi-vs-epsilon}
	\hat{\pi} =  1 - (1 + 2 \hat{\varepsilon})^{-1/2} .
\end{equation}
In the linear approximation used for small deformations, the tensors $\hat{\pi}$ and $\hat{\varepsilon}$ coincide.

Using the deformation tensor defined in this way, we can rewrite (\ref{eq:kdeform}) as
\begin{equation}
	\label{eq:kdeform2}
	\mathbf{k}^\prime =  (1 - \hat{\pi}) \cdot \mathbf{k}
\end{equation}
or
\begin{equation}
	\label{eq:Dk}
	\Delta\mathbf{k} = \mathbf{k}^\prime - \mathbf{k} = - \hat{\pi} \cdot \mathbf{k} .
\end{equation}
The last equation allows us to write the elastic deformation energy (\ref{eq:phidef3}) as a quadratic form in the components of the tensor $\hat{\pi}$:
\begin{equation}
	\label{eq:phidef-Hooke}
	\varphi_\mathrm{def} = \tfrac12 \lambda_{\alpha\beta\gamma\delta} \pi_{\alpha\beta} \pi_{\gamma\delta} ,
\end{equation}
where
\begin{equation}
	\label{eq:lambda}
	\lambda_{\alpha\beta\gamma\delta} = \left< \sum_{\mathbf{k}} \frac{\partial^2 \left<\varphi_\mathrm{grad}\right>}{\partial k_\alpha \partial k_\gamma} k_\beta k_\delta \right>_{[[V^2]^2]}
\end{equation}
is the elastic modulus tensor. Here the angle brackets mean that the tensor in them must be symmetrized under permutations of indices:
\[
[[V^2]^2] : \phantom{x} \lambda_{1234} = \lambda_{2134} =\lambda_{1243} = \lambda_{3412} .
\]
In crystals with cubic symmetry, the elastic modulus tensor has three independent components:
\begin{equation}
	\label{eq:phidef-Hooke2}
	\begin{array}{ll}
		\varphi_\mathrm{def} & = \tfrac12 \lambda_{xxxx} (\pi_{xx}^2 + \pi_{yy}^2 + \pi_{zz}^2) \vspace{0.2cm} \\
		& + \lambda_{xxyy} (\pi_{xx} \pi_{yy} + \pi_{yy} \pi_{zz} + \pi_{zz} \pi_{xx}) \vspace{0.2cm} \\
		& + 2 \lambda_{xyxy} (\pi_{xy}^2 + \pi_{yz}^2 + \pi_{zx}^2) .
	\end{array}
\end{equation}

\section{Reciprocal lattice parameter of a blue phase}
\label{sec:b}

As already mentioned, free energy is a quadratic form in the components of the deformation tensor only if the system is at a point of elastic equilibrium. The equilibrium condition, in particular, determines the size of the unit cell. Note that the period of the structure affects only the gradient part of the free energy, which includes the spatial derivatives of the tensor field $\hat{\chi}(\mathbf{r})$. Indeed, the expression for $\left<\varphi_\mathrm{grad}\right>$ includes reciprocal lattice vectors
\begin{equation}
	\label{eq:k}
	\mathbf{k} = \frac{2\pi}a (hk\ell) = b |hk\ell| \mathbf{n}_\mathbf{k} ,
\end{equation}
where $a$ is the period of the cubic lattice of the crystal; $hk\ell$ are the Miller indices of the Fourier harmonic,
\[
|hk\ell| = \sqrt{h^2 + k^2 + \ell^2} ,
\]
$b=2\pi/a$ is the reciprocal lattice parameter. In addition to the vectors $\mathbf{k}$, the expression (\ref{eq:phigrad-fourier}) includes the Hermitian tensor $\hat{\chi}_\mathbf{k} \cdot \hat{\chi}_\mathbf{k}^\ast$, which for the helical harmonic (\ref{eq:chiLR}) with the coefficient $A=\chi_{\left<hk\ell\right>}$ has the following form:
\begin{equation}
	\label{eq:chi-chi}
	\hat{\chi}_\mathbf{k} \cdot \hat{\chi}_\mathbf{k}^\ast = \tfrac12 \chi_{\left<hk\ell\right>}^2 (\mathbf{m}_1 \mp i \mathbf{m}_2) \otimes (\mathbf{m}_1 \pm i \mathbf{m}_2) .
\end{equation}
Let us substitute (\ref{eq:k}), (\ref{eq:chi-chi}) into the expression (\ref{eq:phigrad-fourier}) and calculate the average density of the gradint free energy as a function of $b$:
\begin{equation}
	\label{eq:phigrad-b}
	\left<\varphi_\mathrm{grad}\right> = \kappa^2 \sum_\mathbf{k} (b^2 |hk\ell|^2 - 2 b |hk\ell| + 1) \lVert \hat{\chi}_\mathbf{k} \rVert^2 .
\end{equation}
By minimizing (\ref{eq:phigrad-b}), we obtain the equilibrium value of the reciprocal lattice parameter:
\begin{equation}
	\label{eq:b}
	b = \sum_\mathbf{k} |hk\ell| \lVert \hat{\chi}_\mathbf{k} \rVert^2 \Big/ \sum_\mathbf{k} |hk\ell|^2 \lVert \hat{\chi}_\mathbf{k} \rVert^2 .
\end{equation}
This expression can be written in a simpler form by introducing the concept of the mean value of an arbitrary quantity $A_\mathbf{k}$, depending on the reciprocal lattice vector:
\begin{equation}
	\label{eq:meanAk}
	\left< A_\mathbf{k} \right> = \frac{1}{X^2} \sum_\mathbf{k} A_\mathbf{k} \lVert \hat{\chi}_\mathbf{k} \rVert^2 .
\end{equation}
Here $X^2$ is the square of the norm of the order parameter, averaged over the unit cell,
\begin{equation}
	\label{eq:meanchi2}
	X^2 = \left<\lVert \hat{\chi}(\mathbf{r}) \rVert^2\right> = \sum_\mathbf{k} \lVert \hat{\chi}_\mathbf{k} \rVert^2 .
\end{equation}
Now the free energy of the gradient (\ref{eq:phigrad-b}) can be written in a more transparent form:
\begin{equation}
	\label{eq:phigrad-k}
	\left<\varphi_\mathrm{grad}\right> = \kappa^2 \left< (k - 1)^2 \right> X^2 ,
\end{equation}
from which it becomes obvious that the reciprocal lattice parameter $b$ in equilibrium minimizes the root mean square deviation of the wavenumber $k$ from unity. Moreover, the parameter $b$ can be written as
\begin{equation}
	\label{eq:b2}
	b = \frac{\left< |hk\ell| \right>}{\left< |hk\ell|^2 \right>} .
\end{equation}
This equation allows us to easily calculate $b$ for the blue phase with all equivalent Fourier harmonics. For example, taking into account that $|110|=\sqrt 2$, for the $O^5$ phase we obtain:
\begin{equation}
	\label{eq:bO5}
	b(O^5) = 1 / \sqrt2 ,
\end{equation}
which corresponds to the wavenumber of helices $k=1$. To find the parameter $b$ for the blue phases $O^8$ and $O^2$, we partially sum the numerator and denominator in the expression (\ref{eq:b}), bringing it to the following form:
\begin{equation}
	\label{eq:b3}
	b = \frac{\sum_{\left<hk\ell\right>} |hk\ell| N_{\left<hk\ell\right>} \chi_{\left<hk\ell\right>}^2}{\sum_{\left<hk\ell\right>}  |hk\ell|^2 N_{\left<hk\ell\right>} \chi_{\left<hk\ell\right>}^2} .
\end{equation}
Here the summation is carried out over sets of crystallographically equivalent vectors $\left<hk\ell\right>$, $N_{\left<hk\ell\right>}$ is the number of wave vectors included in each set. Taking into account that
\begin{equation}
	\label{eq:Nhkl}
	\begin{array}{lll}
		|100| = 1 , & & N_{\left<100\right>} = 6 , \\
		|200| = 2 , & & N_{\left<200\right>} = 6 , \\
		|110| = \sqrt2 , & & N_{\left<110\right>} = 12 , \\
		|211| = \sqrt6 , & & N_{\left<211\right>} =24 ,
	\end{array}
\end{equation}
we obtain
\begin{equation}
	\label{eq:bO8}
	b(O^8) = \frac{\chi_{\left<200\right>}^2 + \sqrt2 \chi_{\left<110\right>}^2 + 2 \sqrt6 \chi_{\left<211\right>}^2}{2 [\chi_{\left<200\right>}^2 + \chi_{\left<110\right>}^2 + 6 \chi_{\left<211\right>}^2]} ,
\end{equation}
\begin{equation}
	\label{eq:bO2}
	b(O^2) = \frac{\chi_{\left<100\right>}^2 + 2 \sqrt2 \chi_{\left<110\right>}^2}{\chi_{\left<100\right>}^2 + 4 \chi_{\left<110\right>}^2} .
\end{equation}

Finally, we note a useful equality that is a consequence of (\ref{eq:k}) and (\ref{eq:b2}):
\begin{equation}
	\label{eq:k2=k}
	\langle k^2 \rangle = \left< k \right> = \frac{\left< |hk\ell| \right>^2}{\left< |hk\ell|^2 \right>} \leqslant 1 .
\end{equation}

\section{One-constant approximation. Bulk modulus of elasticity}
\label{sec:bulk-mod}

To begin with, we restrict ourselves to calculating the elastic moduli in the one-constant approximation ($\eta=1$). Note that the term proportional to $(\eta-1)$ in (\ref{eq:phigrad-fourier}) does not contribute to the average free energy density of the undistorted blue phase, since for the biaxial helices described by the expressions (\ref{eq:expikr}), (\ref{eq:chiLR})
\begin{equation}
	\label{eq:k.chi=0}
	\mathbf{k} \cdot \hat{\chi}_\mathbf{k} = \hat{\chi}_\mathbf{k}^\ast \cdot \mathbf{k} =0 .
\end{equation}
Since under uniform tension/compression the reciprocal lattice vectors decrease/increase without changing direction, the equations (\ref{eq:k.chi=0}) also hold for the vectors $\mathbf{k}^\prime$. The latter means that the one-constant approximation allows one to calculate the bulk modulus of elasticity
\begin{equation}
	\label{eq:K}
	K = \tfrac13 (\lambda_{xxxx} + 2 \lambda_{xxyy}) .
\end{equation}

The energy of elastic deformation is equal to the difference in the energies of the distorted and original structures:
\begin{equation}
	\label{eq:phidef-eta=1}
	\begin{array}{lll}
		\varphi_\mathrm{def} & = & \langle \varphi_\mathrm{grad}^\prime \rangle - \langle \varphi_\mathrm{grad} \rangle \vspace{0.2cm} \\
		& = & \kappa^2 \sum\limits_\mathbf{k} \{ (k^{\prime2} - k^2) \, \mathrm{Tr}(\hat{\chi}_\mathbf{k} \cdot \hat{\chi}^\ast_\mathbf{k}) \vspace{0.2cm} \\ 
		& & \pm 2 i \, \mathrm{Tr}((\mathbf{k}^\prime - \mathbf{k}) \times (\hat{\chi}_\mathbf{k} \cdot \hat{\chi}^\ast_\mathbf{k})) \} ,
	\end{array}
\end{equation}
or
\begin{equation}
	\label{eq:phidef-eta=1b}
	\varphi_\mathrm{def} = \kappa^2 \sum\limits_\mathbf{k}  \{ (k^{\prime2} - k^2)  - 2 (\mathbf{k}^\prime - \mathbf{k}) \cdot \mathbf{n}_\mathbf{k} \} \lVert \hat{\chi}_\mathbf{k} \rVert^2 .
\end{equation}
Here we have taken into account the equation (\ref{eq:chinorm}) and the fact that for the helical Fourier harmonic (\ref{eq:chiLR})
\[
\mathrm{Tr}((\mathbf{k}^\prime - \mathbf{k}) \times (\hat{\chi}_\mathbf{k} \cdot \hat{\chi}^\ast_\mathbf{k})) = \pm i (\mathbf{k}^\prime - \mathbf{k}) \cdot \mathbf{n}_\mathbf{k} \lVert \hat{\chi}_\mathbf{k} \rVert^2 .
\]
Using the equation (\ref{eq:kdeform2}), we obtain
\begin{equation}
	\label{eq:phidef-eta=1c}
	\varphi_\mathrm{def} = \kappa^2 \sum\limits_\mathbf{k}  \{ \mathbf{k} \cdot (\hat{\pi}^2 - 2 \hat{\pi}) \cdot \mathbf{k} + \mathbf{k} \cdot 2\hat{\pi} \cdot \mathbf{n}_\mathbf{k} \} \lVert \hat{\chi}_\mathbf{k} \rVert^2 .
\end{equation}
Further transformations are easily carried out using the cubic symmetry of the undistorted blue phase. Indeed, if on the right side of the equation (\ref{eq:phidef-eta=1c}) we sum over the reciprocal lattice vectors related by crystallographic symmetry, then only the vectors $\mathbf{k}$ and $\mathbf{n}_\mathbf{k}$ change (rotate) in the terms, while the square of the norm $\lVert \hat{\chi}_\mathbf{k} \rVert^2$ and the tensor $\hat{\pi}$ remain constant. Averaging over the cubic group $O$ of the product of a pair of varying vectors gives
\[
\left< \mathbf{k} \otimes \mathbf{k} \right>_O = \tfrac13 k^2 \hat{\delta} , \phantom{x} \left< \mathbf{k} \otimes \mathbf{n}_\mathbf{k} \right>_O = \tfrac13 k \hat{\delta} .
\]
After averaging, the elastic deformation energy takes the form
\begin{equation}
	\label{eq:phidef-eta=1d}
	\varphi_\mathrm{def} = \tfrac13 \kappa^2 \sum\limits_\mathbf{k} \left\{ k^2 \, \mathrm{Tr}(\hat{\pi}^2 - 2 \hat{\pi}) + 2 k \, \mathrm{Tr}(\hat{\pi}) \right\} \lVert \hat{\chi}_\mathbf{k} \rVert^2 ,
\end{equation}
or, taking into account the equation (\ref{eq:k2=k}),
\begin{equation}
	\label{eq:phidef-eta=1e}
	\varphi_\mathrm{def} = \tfrac13 \kappa^2 \left< k^2 \right> X^2 \, \mathrm{Tr}(\hat{\pi}^2)  .
\end{equation}
The last expression corresponds to the isotropic elasticity with Lam\'e parameters
\begin{equation}
	\label{eq:Lame}
	\begin{array}c
		\mu_{\mathrm{L}} = \tfrac12 \lambda_{xxxx} = \lambda_{xyxy} = \tfrac13 \kappa^2 \left< k^2 \right> X^2 , \vspace{0.2cm} \\
		\lambda_{\mathrm{L}} = \lambda_{xxyy} = 0 .
	\end{array}
\end{equation}
The bulk modulus
\begin{equation}
	\label{eq:K2}
	K = \lambda_{\mathrm{L}} + \frac23 \mu_{\mathrm{L}}  = 
	\displaystyle \frac29 \kappa^2 \left< k^2 \right> X^2 .
\end{equation}
Using the equations (\ref{eq:Nhkl}), we find expressions for the modulus $K$ of the blue phases $O^5$, $O^8$, $O^2$:
\begin{equation}
	\label{eq:KO5}
	K(O^5) = \frac83 \kappa^2 \chi_{\langle 110 \rangle}^2 ,
\end{equation}
\begin{equation}
	\label{eq:KO8}
	K(O^8) = \frac43 \kappa^2 \frac{\left( \chi_{\langle 200 \rangle}^2 + \sqrt2 \chi_{\langle 110 \rangle}^2 + 2 \sqrt6 \chi_{\langle 211 \rangle}^2 \right)^2}{\chi_{\langle 200 \rangle}^2 + \chi_{\langle 110 \rangle}^2 + 6 \chi_{\langle 211 \rangle}^2} ,
\end{equation}
\begin{equation}
	\label{eq:KO2}
	K(O^2) = \frac43 \kappa^2 \frac{\left( \chi_{\langle 100 \rangle}^2 + 2 \sqrt2 \chi_{\langle 110 \rangle}^2 \right)^2}{\chi_{\langle 100 \rangle}^2 + 4 \chi_{\langle 110 \rangle}^2} .
\end{equation}

Finally, we express the components of the elastic modulus tensor $\hat{\lambda}$ through the bulk modulus in the one-constant approximation:
\begin{equation}
	\label{eq:lambda-K}
	\lambda_{xxxx} = 3 K , \phantom{x} \lambda_{xxyy} = 0 , \phantom{x} \lambda_{xyxy} = 3 K / 2 .
\end{equation}

\section{The case of two constants. Young's modulus}
\label{seq:Young}

If $\eta\neq1$, then an additional contribution to the elastic deformation energy appears:
\begin{equation}
	\label{eq:phidef-eta} 
	\varphi_\mathrm{def}^{(\eta)} = \kappa^2 (\eta - 1) \sum_{\mathbf{k}} \mathbf{k} \cdot \hat{\pi} \cdot \hat{\chi}_{\mathbf{k}} \cdot \hat{\chi}_{\mathbf{k}}^\ast \cdot \hat{\pi} \cdot \mathbf{k} .
\end{equation}
Here, when summing over the wave vectors $\mathbf{k}$ related by symmetry transformations, the Hermitian tensor $\hat{\chi}_{\mathbf{k}} \cdot \hat{\chi}_{\mathbf{k}}^\ast$ also changes:
\[
\hat{\chi}_{\underline{\mathbf{k}}} \cdot \hat{\chi}_{\underline{\mathbf{k}}}^\ast = R \cdot \hat{\chi}_{\mathbf{k}} \cdot \hat{\chi}_{\mathbf{k}}^\ast \cdot R^{-1} .
\]
Averaging over a cubic point group gives
\begin{equation}
	\label{eq:mean} 
	\begin{array}c
		\langle \mathbf{k} \cdot \hat{\pi} \cdot \hat{\chi}_{\mathbf{k}} \cdot \hat{\chi}_{\mathbf{k}}^\ast \cdot \hat{\pi} \cdot \mathbf{k} \rangle_O = \tfrac16 k^2 \lVert \hat{\chi}_{\mathbf{k}} \rVert^2 \vspace{0.2cm} \\
		\times \{ (\pi_{xx}^2 + \pi_{yy}^2 + \pi_{zz}^2) (1 - \mathfrak{N}_\mathbf{k}) \vspace{0.2cm} \\
		- (\pi_{xx}\pi_{yy} + \pi_{yy}\pi_{zz} + \pi_{zz}\pi_{xx}) (1 - \mathfrak{N}_\mathbf{k}) \vspace{0.2cm} \\
		+ 2 (\pi_{xy}^2 + \pi_{yz}^2 + \pi_{zx}^2) \mathfrak{N}_\mathbf{k} \} ,
	\end{array}
\end{equation}
where
\begin{equation}
	\label{eq:Nfrak} 
	\mathfrak{N}_\mathbf{k} = n_{\mathbf{k},x}^4 + n_{\mathbf{k},y}^4 + n_{\mathbf{k},z}^4 = \frac{h^4 + k^4 + \ell^4}{(h^2 + k^2 + \ell^2)^2}.
\end{equation}
Summing the terms containing $\mathfrak{N}_\mathbf{k}$ gives
\begin{equation}
	\label{eq:sumNfrak} 
	\sum_{\mathbf{k}} \mathfrak{N}_\mathbf{k} k^2 \lVert \hat{\chi}_{\mathbf{k}} \rVert^2 = \mathfrak{A} \left< k^2 \right> X^2 ,
\end{equation}
where
\begin{equation}
	\label{eq:Afrak} 
	\mathfrak{A} = \left< \frac{h^4 + k^4 + \ell^4}{h^2 + k^2 + \ell^2} \right> \Big/ \left< h^2 + k^2 + \ell^2 \right> \leqslant 1.
\end{equation}
Calculating $\mathfrak{A}$ for the blue phases $O^5$, $O^8$, $O^2$ leads to the following expressions:
\begin{equation}
	\label{eq:AfrakO5}
	\mathfrak{A}(O^5) = \frac12 ,
\end{equation}
\begin{equation}
	\label{eq:AfrakO8}
	\mathfrak{A}(O^8) = \frac12 \frac{2 \chi_{\langle 200 \rangle}^2 + \chi_{\langle 110 \rangle}^2 + 6 \chi_{\langle 211 \rangle}^2}{\chi_{\langle 200 \rangle}^2 + \chi_{\langle 110 \rangle}^2 + 6 \chi_{\langle 211 \rangle}^2} ,
\end{equation}
\begin{equation}
	\label{eq:AfrakO2}
	\mathfrak{A}(O^2) = \frac{\chi_{\langle 100 \rangle}^2 + 2 \chi_{\langle 110 \rangle}^2}{\chi_{\langle 100 \rangle}^2 + 4 \chi_{\langle 110 \rangle}^2} .
\end{equation}
As a result, after summing (\ref{eq:phidef-eta}), we obtain the values of the tensor components $\hat{\lambda}$ in the general case:
\begin{equation}
	\label{eq:lambda2}
	\begin{array}l
		\lambda_{xxxx} = 3 K [1 + \tfrac12 (\eta-1) (1 - \mathfrak{A})] , \vspace{0.2cm} \\
		\lambda_{xxyy} = -\tfrac34 K (\eta-1) (1 - \mathfrak{A}) , \vspace{0.2cm} \\
		\lambda_{xyxy} = \tfrac32 K [1 + \tfrac12 (\eta-1) \mathfrak{A}] .
	\end{array}
\end{equation}
Positive definiteness of the deformation energy (\ref{eq:phidef-Hooke}) is determined by the eigenvalues of the elastic modulus tensor
\begin{equation}
	\label{eq:conditions}
	\begin{array}l
		\lambda_1 = \lambda_{xxxx} + 2 \lambda_{xxyy} , \vspace{0.2cm} \\
		\lambda_2 = \lambda_{xxxx} - \lambda_{xxyy} , \vspace{0.2cm} \\
		\lambda_3 = \lambda_{xyxy} ,
	\end{array}
\end{equation}
which are positive for $K,\phantom{;}\eta>0$. The eigenvalues of $\hat{\lambda}$ are related to the bulk and shear moduli by simple relations \cite{Oswald2005}:
\[
K = \lambda_1 / 3 , \phantom{x} G_1 = \lambda_2 / 2 , \phantom{x} G_2 = \lambda_3 .
\]

Let us also calculate Young's modulus, which in the case of an anisotropic medium depends on the direction of tension (compression). In particular, in the case of cubic anisotropy for tension (compression) in the direction of the unit vector $\mathbf{n}$, we have
\begin{equation}
	\label{eq:En}
	\begin{array}{ll}
		E(\mathbf{n}) = & \displaystyle \left[ \frac1{3\lambda_1} - \frac1{3\lambda_2} + \frac1{2\lambda_3} \right. \vspace{0.2cm} \\
		& \left. \displaystyle + \left( \frac1{\lambda_2} - \frac1{2\lambda_3} \right) \left( n_x^4 + n_y^4 + n_z^4 \right) \right]^{-1} .
	\end{array}
\end{equation}
The extreme values of Young's modulus correspond to the crystallographic directions $\langle100\rangle$ and $\langle111\rangle$,
\begin{equation}
	\label{eq:E100}
	E_{\langle 100 \rangle} = \frac{3 \lambda_1 \lambda_2}{2 \lambda_1 +  \lambda_2} = 3K \frac{1 + \tfrac34 (\eta - 1)(1 - \mathfrak{A})}{1 + \tfrac14 (\eta - 1)(1 - \mathfrak{A})} ,	
\end{equation}
\begin{equation}
	\label{eq:E111}
	E_{\langle 111 \rangle} = \frac{3 \lambda_1 \lambda_3}{\lambda_1 +  \lambda_3} = 3K \frac{1 + \tfrac12 (\eta - 1) \mathfrak{A}}{1 + \tfrac16 (\eta - 1) \mathfrak{A}} .
\end{equation}

\section{Discussion}
\label{sec:discussion}

Overall, it can be concluded that liquid crystals with low chirality ($\kappa\ll1$) exhibit similar  behavior of elastic moduli in all blue phases. Let us examine this in more detail.

The gradient contribution to the free energy turns out to be small compared to the bulk one,
\[
\langle \varphi_\mathrm{grad} \rangle \ll \langle \varphi_\mathrm{bulk} \rangle ,
\]
and therefore it can be assumed that the coefficients $\chi_{\langle hk\ell\rangle}$ for each phase are determined from the condition of minimum bulk energy. The expression (\ref{eq:phibulk-fourier}) for $\langle \varphi_\mathrm{bulk} \rangle$ contains no parameters other than temperature. Consequently, the coefficients $\chi_{\langle hk\ell\rangle}$ are also functions of temperature only and do not depend on the parameters $\kappa$ and $\eta$ of the gradient energy. The temperature dependence has a power-law form,
\begin{equation}
	\label{eq:temp1}
	\chi_{\langle hk\ell \rangle} \propto (\tau_\mathrm{m} - \tau)^{1/2} ,
\end{equation}
where $\tau_\mathrm{m}$ is the temperature at which the local minimum of free energy corresponding to this blue phase first appears. Given that blue phases exist in a very narrow temperature range between the isotropic and helical phases, we can neglect the relative changes in the coefficients $\chi_{\langle hk\ell\rangle}$ and introduce a single order parameter $X$:
\begin{equation}
	\label{eq:OP}
	X^2 = \left< \lVert \hat{\chi}(\mathbf{r}) \rVert^2 \right> = \sum_{\left<hk\ell\right>} N_{\left<hk\ell\right>} \chi_{\left<hk\ell\right>}^2 .
\end{equation}
All elastic moduli are proportional to $X^2$ and, therefore, change with temperature according to a linear law:
\begin{equation}
	\label{eq:temp2}
	\hat{\lambda} \propto (\tau_\mathrm{m} - \tau) .
\end{equation}
In this case, each blue phase has its own temperature $\tau_\mathrm{m}$ and proportionality coefficient.

\begin{figure}[h]
	\begin{center}
		\includegraphics[width=8cm]{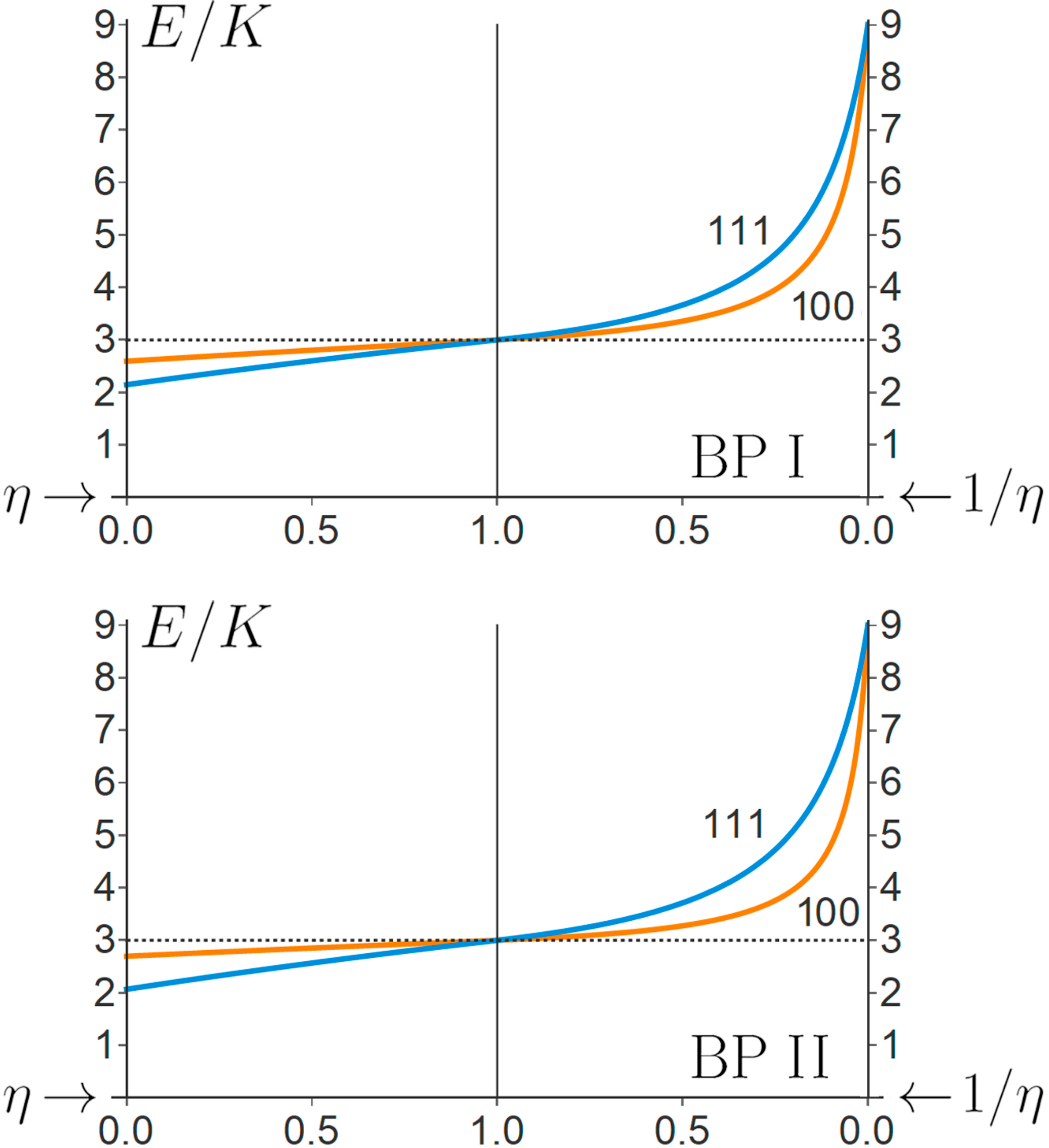}
		\caption{\label{fig:EvsH} Dependence of the ratios of Young's moduli $E_{\langle100\rangle}$ and $E_{\langle111\rangle}$ to the bulk modulus $K$ on the parameter $\eta$ for the blue phases $O^8$ (BPI) and $O^2$ (BPII) in the rigid tensor approximation.}
\end{center}
\end{figure}

Next, we will consider the characteristic features of the elastic properties of the blue phases that are not related to temperature. To do this, we divide all the elastic moduli by the bulk modulus $K$, thereby eliminating the temperature dependence. The tensor $\hat{\lambda}/K$ depends on two quantities: the constant $\eta=K_0/K_1$, which characterizes the physical system, and $\mathfrak{A}$, a number associated with the blue phase. To calculate $\mathfrak{A}$, it is sufficient to know the ratios of the coefficients $\chi_{\langle hk\ell\rangle}$. For example, using the data from \cite{Grebel1983,Belyakov1985} one can find the ratio
\begin{equation}
	\label{eq:O8chi}
	\chi_{\langle 200 \rangle} : \chi_{\langle 110 \rangle} : \chi_{\langle 211 \rangle} = 1 : 0.934 : -0.329
\end{equation}
for the $O^8$ phase, and
\begin{equation}
	\label{eq:O2chi}
	\chi_{\langle 100 \rangle} : \chi_{\langle 110 \rangle} = 1 : -0.396 
\end{equation}
for the $O^2$ phase. Substituting these ratios into the equations (\ref{eq:AfrakO8}), (\ref{eq:AfrakO2}), we find
\begin{equation}
	\label{eq:AfrakO8O2}
	\mathfrak{A}(O^8) = 0.698, \phantom{xx} \mathfrak{A}(O^2) = 0.807 .
\end{equation}
Knowing these values, we can plot the dependences of $E_{\langle100\rangle}/K$, $E_{\langle111\rangle}/K$ on $\eta$ for the blue phases BPI and BPII (Fig.~\ref{fig:EvsH}). The figures show that the $E/K$ functions increase monotonically in the range $\eta\in[0,\infty)$. In the one-constant approximation ($\eta=1$), blue phase lattices are elastically isotropic, with Young's modulus equal to three times the bulk modulus. If $\eta\neq 1$, then cubic anisotropy of elastic properties arises in accordance with the expression (\ref{eq:En}). In particular,
\begin{equation}
	\label{eq:etaYoung}
	\begin{array}l
		E_{\langle100\rangle} > E_{\langle111\rangle} , \phantom{x} \eta < 1 , \vspace{0.2cm} \\
		E_{\langle100\rangle} < E_{\langle111\rangle} , \phantom{x} \eta > 1 .
	\end{array}
\end{equation}

The Poisson's ratio $\nu$ is zero in the one-constant approximation ($\eta=1$). This means that when the lattice is stretched (compressed) in any direction, its transverse dimensions do not change. When $\eta\neq1$, the Poisson's ratio, like Young's modulus, depends on the direction. However, the sign of $\nu$ correlates with the parameter $\eta$:
\begin{equation}
\label{eq:etaPoisson}
\mathrm{sgn}[\nu (1 - \eta)] = +1 . 
\end{equation}
In particular, for $\eta>1$, the lattices of blue phases behave as {\it auxetics}, i.e., materials with negative Poisson's ratio, $\nu<0$. This property can be possessed by artificially created materials \cite{Ren2018}, but auxetics are also found among ordinary crystals, including those with cubic symmetry \cite{Goldstein2013}. However, in the case of blue phases, there is an important difference. Since the number of molecules in the unit cell is not fixed, an increase in its size does not automatically lead to a decrease in the density. Here one can see an analogy between the movements of the boundary of the unit cell of the blue phase and the domain wall in ferroelectrics and ferromagnets. Note that in nematic and cholesteric LCs, the Frank elastic modulus $K_{22}$ is usually significantly smaller than $K_{11}$ and $K_{33}$, and, consequently, $\eta>1$. With a typical value of $\eta\approx3$ \cite{Wright1989}, the Poisson's ratio of the blue phases BPI and BPII is negative and varies approximately in the range from $-0.1$ to $-0.2$ depending on the direction of tension.

It is worth separately considering the behavior of elastic properties at $\eta\rightarrow\infty$ ($K_1\rightarrow0$). Figure~\ref{fig:EvsH} shows that at $\eta=\infty$ Young's modulus does not depend on the direction of tension (compression). In this case $E=9K$, and if we look at the expressions for the Lam\'e parameters,
\begin{equation}
\label{eq:Lame2}
\lambda_\mathrm{L} = \frac{3K (3K - E)}{9K - E} , \phantom{x} \mu_\mathrm{L} = \frac{3KE}{9K - E} ,
\end{equation}
this may seem like catastrophic behavior. Here, however, it should be noted that for large values of the parameter $\eta$, the low chirality condition should look as follows:
\begin{equation}
\label{eq:kappa-eta}
\kappa \ll \eta^{-1/2} .
\end{equation}
This means that as $\eta\rightarrow\infty$ the bulk modulus $K\sim\kappa^2$ and Young's modulus $E$ tend to zero. The reason for this is the disappearance of the first term in the gradient energy density $\varphi_\mathrm{grad}$ (\ref{eq:varphi}), which is responsible for the twist of the order parameter field $\hat{\chi}(\mathbf{r})$ (in the Frank--Oseen energy this corresponds to $K_{22}=0$). Thus, the dependence of the free energy on the average wavenumber $\langle k\rangle$ of the helices and, consequently, on the size of the unit cell disappears. When attempting to stretch (compress) the lattice along a chosen direction, the cell experiences uniform tension (compression): Poisson's ratio $\nu=-1$. In this case, the system retains anisotropic elasticity, characterized by the shear moduli
\begin{equation}
\label{eq:shear}
\begin{array}l
	G_1 = \tfrac14 (1 - \mathfrak{A}) \kappa^2 \eta \langle k^2 \rangle X^2 , \vspace{0.2cm} \\
	G_2 = \tfrac16 \mathfrak{A} \kappa^2 \eta \langle k^2 \rangle X^2 .
\end{array}
\end{equation}

The theoretical study of the elastic moduli of blue phases of cholesteric liquid crystals with low chirality, carried out in this work, can be considered a necessary step for further consideration of other physical properties: elastic-optical, electro-optical, elastic-electrical, which are important for practical applications. All these properties can be investigated in a unified manner within the framework of the Landau--de~Gennes theory used here.

\section*{Acknowledgements}

The authors are grateful to M.~V.~Gorkunov, E.~I.~Demikhov, P.~V.~Dolganov, V.~K.~Dolganov, and E.~I.~Kats for their critical remarks and useful discussion, as well as to A.~V.~Mamonova for her assistance with calculations.

\section*{Funding}

This work was supported by the Russian Science Foundation, grant No. 23--12--00200.

%\newpage

\end{document}